\documentclass[10pt,letterpaper]{article}
\usepackage{opex3}
\usepackage{amsmath}
\begin{document}

%%%%%%%%%%%%%%%%%% title page information %%%%%%%%%%%%%%%%%%
\title{Pendell\"{o}sung effect in photonic crystals}

\author{S. Savo$^1$, E. Di Gennaro$^1$, C. Miletto$^1$,
A. Andreone$^1$, P. Dardano$^2$, L. Moretti$^2,3$, V. Mocella$^2$}

\address{$^1$CNISM and Department of Physics, Universit$\grave{a}$ di
Napoli ``Federico II'', Piazzale Tecchio 80, I-80125 Naples,ITALY
\\$^2$IMM - CNR, Sezione di Napoli, via P. Castellino 111,I-80131 Naples, ITALY
\\$^3$DIMET - University "Mediterranea" of Reggio Calabria
Località Feo di Vito, \\I-89100 Reggio Calabria (Italy)}

\email  {emiliano.digennaro@na.infn.it}

\begin{abstract}
At the exit surface of a photonic crystal, the intensity of the
diffracted wave can be periodically modulated, showing a maximum in
the "positive" (forward diffracted) or in the "negative"
(diffracted) direction, depending on the slab thickness. This
thickness dependence is a direct result of the so-called
Pendell\"{o}sung phenomenon, consisting of the periodic exchange
inside the crystal of the energy between direct and diffracted
beams. We report the experimental observation of this effect in the
microwave region at about $14 GHz$ by irradiating 2D photonic
crystal slabs of different thickness and detecting the intensity
distribution of the electromagnetic field at the exit surface and
inside the crystal itself.
\end{abstract}

 \ocis{(050.1960) Diffraction theory; (260.2110) Electromagnetic
theory; (290.4210) Multiple scattering; (999.9999) Photonic crystal}

%%%%%%%%%%%%%%%%%%%%%%% References %%%%%%%%%%%%%%%%%%%%%%%%%

%%%%%%%%%%%%%%%%%%%%%%%%%%  body  %%%%%%%%%%%%%%%%%%%%%%%%%%
\section{Introduction}
Since the original formulation of the diffraction theory from Ewald
\cite{PPE1}, the Pendell\"{o}sung effect was predicted as a periodic
exchange of energy between interfering wave-fields. The German term
comes from the formal analogy between the mechanical system composed
by coupled pendula and the optical problem, where many waves
contribute to the optical field.  In this formal analogy pendulum is
the counterpart of wave whereas the temporal dependence of the
mechanical problem corresponds to the spatial dependence in the
considered optical problem \cite{PPE2}. Pendell\"{o}sung is a
relatively well known effect of Dynamical Diffraction Theory (DDT),
a rigorous formalism accounting for multiple scattering effects that
are especially important in X-ray, electron and neutron diffraction
from perfect crystals \cite{AA}. The requirement of high quality
crystals explains why the first experimental observation of the
Pendell\"{o}sung effect has been obtained in 1959 only in X-ray
measurements \cite{NK}, and some years later in neutron diffraction
\cite{DS,CGS}. Recently, using the coherence of third generation
synchrotron beams, Pendell\"{o}sung fringes produced by a plane wave
exiting a Si crystal have been recorded \cite{VM3}.

Photonic crystals (PhCs) are artificial periodic structures
reproducing natural crystals at different length scale. PhCs can
control and manipulate the flow of light in many different ways,
since they exhibit a variety of properties, spanning from full
photonic band gap to anomalous dispersion phenomena, including
superprism and negative refraction effects. The wide range of
characteristics shown by PhCs gave rise in the last decade to a
multitude of new ideas for optoelectronic integrated devices and
systems. Novel concepts of mirrors, waveguides, resonators, and
frequency converters based on photonic crystals, to mention a few
examples, have been proposed. Indeed, the band theory of the
electrons in solids, that is usually the main reference of the PhC
theory, is strongly inspired by DDT (as extensively discussed in the
pioneering Born's solid state textbook \cite{MB}), that represents
one of the first example of two-state theory - later became very
popular in modern physics - such as up and down spins, electron and
hole pairs, etc.

It is not surprising therefore that the Pendell\"{o}sung effect has
been predicted for photonic crystals too. In 2D case, it has been
thoroughly studied using both analytical and numerical methods as a
function of the PhC contrast index, beam incident angle, and light
polarization \cite{VM1}. Moreover, this study has been extended to
opal 3D photonic crystals, where the dependence of diffraction
intensity as a function of  the layers number has been investigated
using a scattering matrix approach \cite{AB}. On the experimental
side the properties of microwave diffraction in periodic structures
have been reported in literature by measuring the pattern of
backscattered waves in two dimensional artificial dielectric media
\cite{OF}. Recently, the Pendell\"{o}sung effect has been detected
also in the optical regime in volume holographic gratings, observing
the oscillatory behavior of the angular selectivity of the
diffracted light \cite{MLC}.

In this work, we present an accurate theoretical study and precise
measurements of the Pendell\"{o}sung effect in photonic crystals by
illuminating  with a plane waves beam in the microwave region 2D
square lattice PhCs having different number of rows (i.e. slabs with
different thicknesses) and detecting the electromagnetic field
inside and outside each slab. We show that under particular
conditions the intensity of the diffracted wave at the exit surface
can be periodically modulated with the slab thickness, presenting a
maximum in the "positive" (forward diffracted) or in the "negative"
(diffracted) direction. Moreover, we observe that inside the crystal
the energy is periodically exchanged between direct and diffracted
beams.

\section{Theoretical analysis}
The Pendell\"{o}sung effect in PhCs can be understood as a beating
phenomenon due to the phase modulation between coexisting plane wave
components, propagating in the same direction. The coexistence is
possible because such wavevectors are associated to two adjacent
bands that are overlapped, for a given frequency, in correspondence
of suitably chosen PhC parameters.
%%%%%%%%%%%%%%%%%%%%%%%%%%%%
\begin{figure}[htbp]
\centering  \includegraphics[width=11cm]{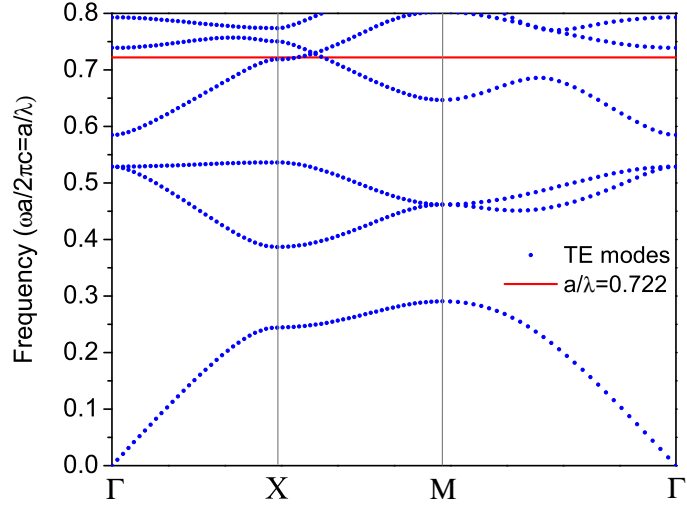}
\caption{\label{Figure0}The band structure of the square-lattice PhC
for the $TE$ polarization. The red line represents the normalized
frequency $\omega_n=0.722$ at which the Pendell\"{o}sung effect
takes place (colour online).}
\end{figure}
%%%%%%%%%%%%%%%%%%%%%%%%%%%%

In our case the 2D PhC consists of dielectric cylinders in air
(dielectric permittivity  $\varepsilon_r= 8.6$) arranged in a square
geometry and having $r/a = 0.255$, where $r$ is the cylinder radius
and $a$ is the lattice constant. If $TE$ polarization (electric
field parallel to the rods axis) is considered, an overlap occurs
between the forth and the fifth mode for a normalized frequency
$\omega_n=\nu a/c = a/\lambda=0.722$, as shown in Fig.\ref{Figure0}.
Moreover, the crystal orientation is fixed such that the normal at
its surface is along the $XM$ direction. Hence all possible
wavevectors excited into the PhC will have the same tangential
component lying on $XM$.

\begin{figure}[htbp]
\centering  \includegraphics[width=11cm]{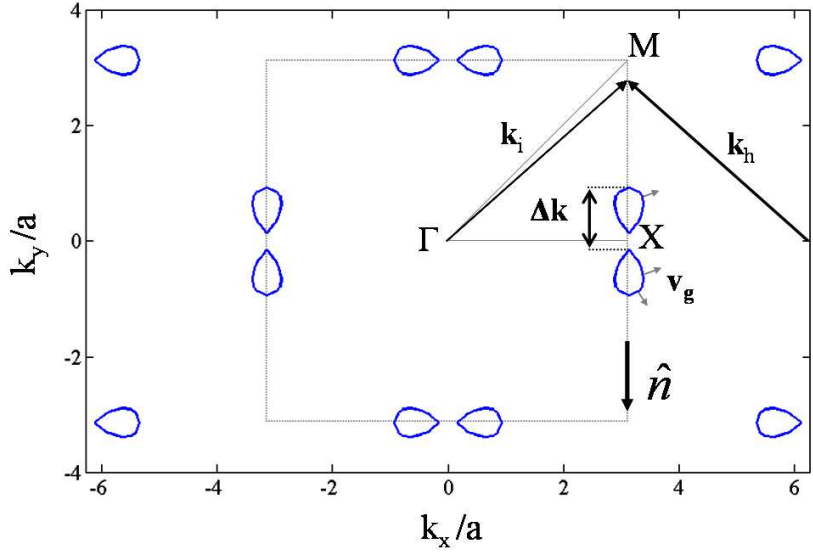}\\
  \caption{\label{Figure1}The reciprocal space with the first
Brillouin zone (dotted line) and symmetry points for the
square-lattice PhC. The contours for the normalized frequency
$\omega_n=0.722$ are plotted. Arrows indicate the directions of
group velocity $\mathbf{v_g}$, whereas $\mathbf{\hat{n}}$ shows the
normal to the incident surface (colour online).}
\end{figure}

The Pendell\"{o}sung phenomenon is analyzed in this context for an
incident wavevector that satisfies the Bragg law \cite{VM1,VM2}. In
Fig.\ref{Figure1} we show in the reciprocal space the first
Brillouin zone and the corresponding symmetry points for the square
lattice PhC under study. Considering the reciprocal lattice vector
that enforces the momentum conservation oriented along $\Gamma X$ in
the first Brillouin zone, the Bragg law is fulfilled when the
projection of the incident wavevector coincides with $\Gamma X$, so
that $\mathbf{k}_{h}$ is the diffracted wavevector whereas
$\mathbf{k}_{i}$ is the incident one. Using the dispersion surfaces
(or Equi-Frequency Surfaces, EFSs), that represent the loci of
propagating wavevectors for a fixed frequency, we are then able to
evaluate the relevant parameters of the beating effect. The
wavevectors inside the PhC are determined by the intersection
between each EFS and the $XM$ direction. Amongst the different
intersections, only wavevectors having group velocity oriented
inside the crystal - in Fig. \ref{Figure1} in opposite direction
respect to the external normal to the incident surface - will be
effectively excited.
%%%%%%%%%%%%%%%%%%%%%%%%%%%%
%%%%%%%%%%%%%%%%%%%%%%%%%%%%

Consider for instance the contribution of the incident wave: there
is an interference between two excited components, with the
respective wavevectors pointing in two different directions. This
produces a spatial periodic modulation along the wavevectors
difference vector ${\Delta \mathbf{k}}$. The modulation distance in
the real space along the PhC normal direction is therefore
$\Lambda_0 = 2\pi/\Delta k$. The same effect occurs also for the
diffracted wave, giving rise to a spatial modulation with the same
length but $180^\circ$ out-of-phase in respect to the previous case.

As a consequence of the Pendell\"{o}sung effect, the intensity $I$
at the exit surface is harmonically modulated as a function of the
thickness $t$ \cite{VM1}. When $t$ is an even multiple of half the
Pendell\"{o}sung distance, the beam at the exit surface is parallel
to the incident beam, forming a positive angle respect to the PhC
normal. On the other hand, when $t$ is an odd multiple of
$\Lambda_0$ the beam at the exit surface is completely directed
along the Bragg diffracted direction, forming a negative angle
respect to the PhC normal. Denoting by + and - the two possible
directions at the exit surface, this is summarized by:
\begin{equation}\label{eq1}
                  \begin{aligned}
                  t&=~ ~2m\frac{\Lambda}{2} &\Rightarrow max(I_+) \\
                 t&=(2m-1)\frac{\Lambda}{2} &\Rightarrow max(I_-)
                  \end{aligned}
\end{equation}
where $m = 1, 2,\ldots$. \\Forcing the Pendell\"{o}sung distance
$\Lambda_0$ be an even number of the lattice constant $a$, eq.
(\ref{eq1}) holds for any number $n$ of PhC rows. In particular,
$\Lambda_0 =4a $ ensures that the intensity maxima of the exit waves
changes periodically if $n$ is even, and that the energy beam
equally splits between positive and negative direction if $n$ is
odd.

From the EFSs analysis, assuming a TE polarization, we found that an
angle $\theta_i= 43.8^\circ$ and a normalized frequency
$\omega_n=0.722$ for the incident wave satisfy both the Bragg law
and the peculiar condition $\Lambda_0= 4a$.

\section{Experimental setup}
The experimental results are obtained on 2D PhCs having a different
number of rows inserted in a waveguide. First, the electromagnetic
wave transmitted by the periodic structure is measured at the exit
of the PhC for different crystal thickness and its spatial
distribution is shown. Then, the periodic modulation of the
intensity of the diffracted waves with respect to $n$ is reported.
Finally, a comparison along selected directions inside the photonic
crystal between the electric field distribution measured and
simulated using a Finite Difference Time Domain (FDTD) method is
presented.

Measurements are carried out by placing alumina rods with nominal
permittivity $\varepsilon_r= 8.6$, radius $r = 0.4 cm$ and height $h
= 1 cm$ in a square geometry with $r/a=0.255$ ($a = 1.57 cm$)
sandwiched in an aluminum parallel-plate waveguide terminated with
microwave absorbers. Since the loss tangent of alumina is extremely
small at the frequency relevant for this work ($\tan
\delta<10^{-4}$), dielectric losses can be neglected. Due to the
presence of metallic plates acting as mirrors, current lines that
are perpendicular to the plates can be considered as infinitely
long, as stated by the well-known mirror theorem. For the same
reason the electric fields produced by these currents are constant
along the same direction and thus the whole system acts as a 2D
structure.

The microwave photonic crystal is built in the shape of a $38.5cm$
wide slab (25 rod columns), with a thickness that can be varied
adding or removing rows. A dipole antenna is used as source,
oriented to produce an electric field parallel to the rods axis and
operating at the frequency of $13.784 GHz$, in order to reproduce
the same normalized frequency $a/\lambda$ of the theoretical model.
Due to the waveguide characteristics, the TEM mode only can
propagate up to $15 GHz$. The maps of the real part of the electric
field are collected by using a HP8720C Vector Network Analyzer and
another dipole antenna as a detector, that moves along the waveguide
plane using an x-y step motor. The thickness dependence has been
investigated based on the observation of the beams at the exit of
the crystal-air interface. We focused our analysis on structures
with a number of rows $n$ ranging from 1 to 10.
%%%%%%%%%%%%%%%%%%%%%%%%%%%%
\begin{figure}[htbp]
\centering \includegraphics[width=11cm]{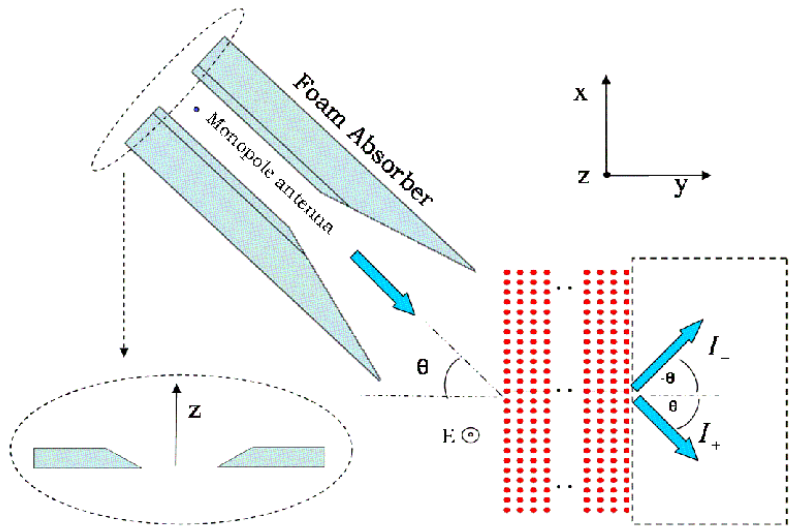}\\
  \caption{\label{Figure2} Schematic layout of the experiment
carried out on the square-lattice PhC slab having 25 rods columns
and a number of rows n varying from 10 to 1. The dashed line box
represents the scanned area during the measurements (colour
online).}
\end{figure}
%%%%%%%%%%%%%%%%%%%%%%%%%%%%

Fig. \ref{Figure2} shows the scheme of the measurement. Particular
attention has been paid to the source characteristics. The incident
beam has to be as collimated and directive as possible, ideally
consisting of a single wavevector only. To realize the experiment,
we inserted in the parallel plate waveguide two parallel microwave
absorber stripes, having the role to "guide" the electromagnetic
wave. The channel is $50 cm$ long and $10 cm$ wide, with tapered
sidewalls in order to ensure a good matching condition at the
air-to-absorber interface. The field generated by the dipole source
is centered into the absorbers channel. To limit the diffraction  at
the exit of the waveguide, a phenomenon that strongly reduces the
beam directivity, the channel section closer to the PhC interface
has been shaped into a triangular profile. All these solutions
provide a beam source having transmission properties close to ideal
ones.

The incoming beam is then oriented at $43.8^\circ$ respect to the
normal to the PhC interface, whereas the two arrows exiting the
surface in both the forward diffracted and diffracted directions
represent the signals transmitted through the wave guide and the
crystal. Furthermore, the angle that describes the outcoming waves
is the same as the source. In the image plane, a tiny dipole antenna
(radius $\sim 0.6 mm$) scans an area $20 cm$ long and $40 cm$ wide
contiguous to the crystal-air interface, in steps of 4 mm in both x
and y direction. As said before, for a fixed frequency and source
orientation the intensities $I_+$ and $I_-$ reach a maximum or a
minimum value depending on the difference between the wavevectors
inside the crystal and, in turn, on the crystal thickness.

\section{Results and discussion}
\begin{figure}[htbp]
\centering \includegraphics[width=11cm]{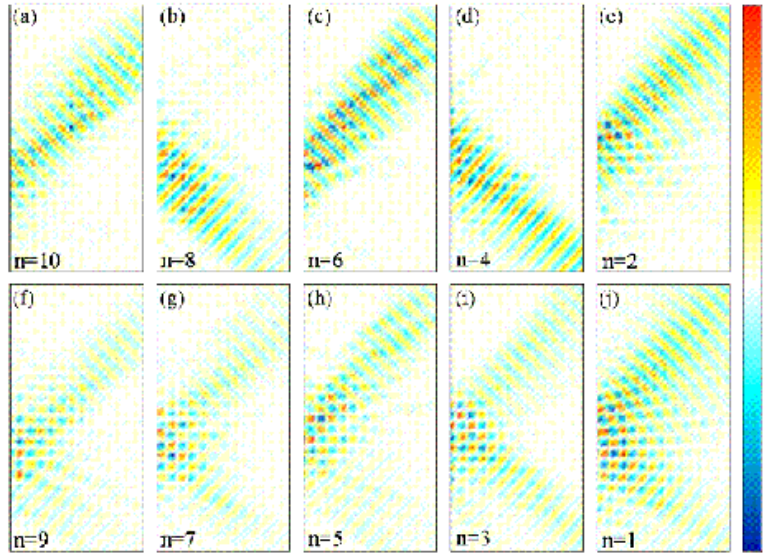}\\
  \caption{\label{Figure3}(a)-(e): mapping of the measured
electric field (real part) in a normalized scale for even n;
(f)-(j): mapping of the measured electric field (real part) in a
normalized scale for odd n (colour online).}
\end{figure}

In Fig. \ref{Figure3} (a)-(e) the real part of the electric field
experimentally detected in different crystal configurations is
mapped in the image plane, using a normalized scale. In Fig.
\ref{Figure3}(a) the  spatial distribution is shown for the case $n
= 10$. The maps for the other cases of crystals with an even number
of rows ($n = 8, 6, 4, 2$) are presented in Figs. \ref{Figure3}(b),
(c), (d), (e),  respectively.
%%%%%%%%%%%%%%%%%%%%%%%%%%%%
%%%%%%%%%%%%%%%%%%%%%%%%%%%%
Starting the data analysis from the crystal consisting of 10 rows,
that is an odd multiple of $\Lambda_0/2$, in this case the beam, as
expected, is fully transmitted in the diffracted direction. On the
contrary, when the PhC consists of 8 rows, the beam exits its
surface in the forward diffracted direction (Fig. \ref{Figure3}(b)).
By reducing the thickness down to 2 rows for any even $n$, the
transmitted beam alternatively bends from the negative to the
positive direction, as shown in Figs. \ref{Figure3}(c)-(e). Other
beams related to higher order of diffraction are negligible.
Measurements clearly show therefore that for an even number of rows
the involved energy is almost entirely concentrated along one exit
direction only.  It is also clear from the images that the
transmitted field propagates along regular and periodic equiphase
planes, in agreement with numerical simulations.

Let us now analyze the experimental results for crystals having an
odd number of rows. In this case, according to the periodical
modulation predicted for the field intensity at the exit surface,
the thickness is such that at the crystal-air interface the
transmitted energy is equally divided in both positive and negative
directions. This is shown in Fig. \ref{Figure3}(f)-(j): the
electromagnetic beam in the image plane actually splits in two rays
having approximately the same intensity, with equiphase planes
clearly evident in both directions. It is worth noting that the case
with  $n = 1$ (Fig. \ref{Figure3}(j)) reduces to the well known
Bragg grating. The fundamental feature of the Pendell\"{o}sung
effect is the spatial periodic modulation of the transmitted field
amplitude with the crystal thickness.
%%%%%%%%%%%%%%%%%%%%%%%%%%%%
\begin{figure}[htbp]
\centering \includegraphics[width=9.5cm]{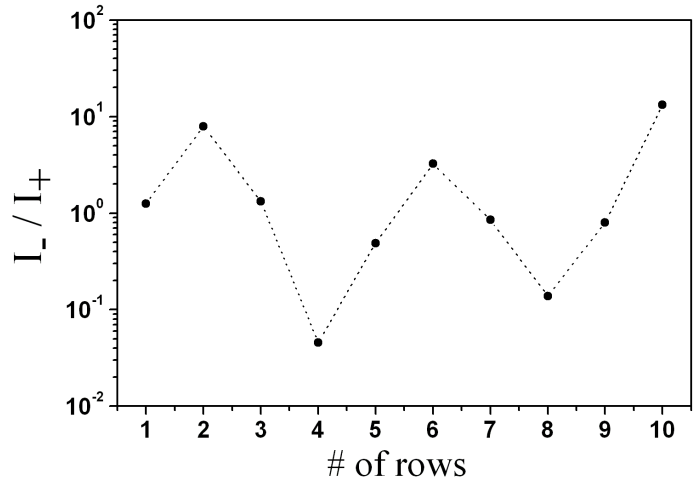}\\
\caption{\label{Figure4}The measured electric field intensity ratio
$I_- / I_+$  for all the crystal configurations considered. The case
of 10 rows corresponds to the maximum thickness
$t=(10a+2r)=16.4cm$.}
\end{figure}
%%%%%%%%%%%%%%%%%%%%%%%%%%%%

We then compared the electric field maximum intensity measured along
the two different (positive and negative) transmitted directions as
a function of the photonic crystal row number $n$ (thickness). As
shown in the Fig. \ref{Figure4} using a semi-log scale, the
intensity ratio $I_- / I_+$ changes periodically, being
approximately equal to 1 for any odd $n$, and exhibiting pronounced
maxima and minima alternatively for any even $n$.

We also evaluated the electromagnetic field distribution inside the
PhC slab. FDTD simulations are performed considering a plane
monochromatic incident wave having a rectangular transverse profile.
The propagation in the slab is of course well different from that in
free-space since Bloch modes will be excited and therefore a strong
modulation of the electromagnetic field is expected. In particular,
in the Pendell\"{o}sung phenomenon, the positively and negatively
refracted components of the incident wave interfere each other
inside the crystal and give rise to a periodic exchange of energy.
This translates in a spatial modulation of the field intensity, as
it can be clearly seen when the dielectric contrast is not very high
\cite{VM2}.

\begin{figure}[htbp]
\centering \includegraphics[width=11cm]{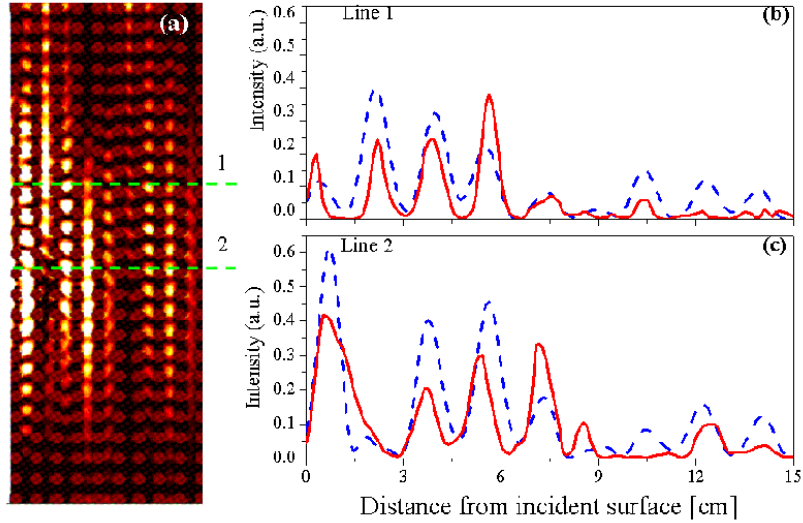}\\
  \caption{\label{Figure5}(a) FDTD simulation of the propagation
pattern inside a crystal consisting of 10 rows of the 13.784 GHz
plane wave modulated by a rectangular profile and incident at an
angle of $43.8^\circ$ across the $XM$ interface; (b) \& (c):
electric field intensity distribution (blue dashed lines) along line
1 and 2 respectively compared with the experimental results (red
solid lines) (colour online).}
\end{figure}

In the case discussed here the strong Bloch modulation makes the
visualization of the wave pattern quite difficult. Moreover, when
the contrast is high, the intensity maxima inside the rods mask the
distribution in the outside region. Therefore, for the sake of
clarity, we have suppressed the field inside the dielectrics.
Results are shown in Fig. \ref{Figure5}(a). It is worth mentioning
that the spatial modulation observed in the crystal does not affect
the energy direction, which remains normal to the PhC interface. To
compare the numerical simulations with experimental data we have
then measured the internal field along selected directions normal to
the PhC interface. Particular attention has been paid to ensure that
the detector antenna moves perfectly parallel to the dielectric
rods.
%%%%%%%%%%%%%%%%%%%%%%%%%%%%
%%%%%%%%%%%%%%%%%%%%%%%%%%%%%

Figures \ref{Figure5}(b) and \ref{Figure5}(c) shows the simulated
longitudinal profile of the field intensity along two different
lines, $(1)$ and $(2)$ respectively, and the corresponding
experimental results properly rescaled. In spite of the strong field
modulation, the presence of peaks and valleys centered in different
positions along the PhC normal direction and corresponding to
different minima and maxima in the wave energy on the two
longitudinal lines is evident, as expected by the theory. Besides
that, the decrease in the intensity as far as the electromagnetic
wave propagates inside the crystal reflects the energy radiated from
the finite-size PhC. This can be also observed in Fig.
\ref{Figure5}(a).

\section{Conclusion}
We have designed and realized an experiment to measure the
Pendell\"{o}sung effect in PhCs, where two beams are $180^\circ$ out
of phase each other, giving rise to the effect of energy flow
swapping back and forth between two different directions into the
crystal. Pendell\"{o}sung can be in principle exploited to design a
new kind of photonic crystal based devices. In a previous paper, it
has been suggested that this effect can be used to realize a passive
polarising beam splitter \cite{VM2}. In the case presented here, we
have shown that adding or removing rows in a finite thickness 2D PhC
can dramatically change the direction and the features of the
electromagnetic wave transmitted through the slab. Data are in a
very good agreement with calculations based on the classical
approach of the Dynamical Diffraction Theory, showing that this
rigorous formalism can be successfully applied to predict some
anomalous features exhibited by PhCs. In particular, we underline
that the Pendell\"{o}sung effect has been extensively used for the
accurate determination of the structure factors in real crystals
\cite{NK2}. It can be extremely useful in photonic crystals too,
allowing precise measurements of some peculiar properties such as
the dielectric contrast.

\section*{Acknowledgement}
This work has been partially funded by MIUR under the PRIN-2006
grant on "Study and realization of metamaterials for electronics and
TLC applications".

\end{document}